\documentclass[preprint,3p]{elsarticle}
\usepackage[english]{babel}
\usepackage[utf8]{inputenc}

\usepackage{color,soul}
\usepackage[english]{babel}
\usepackage{graphicx}
\usepackage{longtable}
\usepackage[utf8]{inputenc}
\usepackage[]{tabularx}
\usepackage{circuitikz}
\usepackage{booktabs,multirow,array}
\usepackage{subfigure}
\usepackage[]{amsmath}

\usepackage{resizegather}
\usepackage{mathtools, cuted}
\usepackage{rotating}
\usepackage{todonotes}
\usepackage{tabularx}

\definecolor{mygreen}{rgb}{0.0,0.7,0.4}

\usetikzlibrary{patterns,decorations.pathmorphing,positioning}
\newcommand{\vc}[1]{\mbox{\boldmath$#1$}}
\newcommand{\matr}[4]{
\left[\begin{array}{cc}
#1 & #2\\
#3 & #4
\end{array}\right]
}

\newcommand{\vect}[2]{
\left[\begin{array}{c}
#1 \\
#2
\end{array}\right]
}

\begin{document}
	\begin{frontmatter}
\title{A new electromechanical analogy approach based on electrostatic coupling for vertical dynamic analysis of planar vehicle models}

%
%
%
\author[esi]{J. López-Martínez}
\ead{javier.lopez@ual.es}
\author[esi]{J. Castillo}
\ead{jcm355@ual.es}
\author[sev]{D. García-Vallejo}
\ead{dgvallejo@us.es}
\author[esi]{A. Alcayde}
\ead{aalcayde@ual.es}
\author[esi]{Francisco G. Montoya}
\ead{pagilm@ual.es}

\address[esi]{CIAIMBITAL Research Centre, ceiA3, Department of Engineering, University of Almería, Carretera de Sacramento s/n, 04120 Almería (Spain)}  
\address[sev]{Department of Mechanical Engineering and Manufacturing, Universidad de Sevilla. Camino de los Descubrimientos~s/n, 41092 Seville (Spain).}


\begin{abstract}
Analogies between mechanical and electrical systems have been developed and applied for almost a century, and they have proved their usefulness in the study of mechanical and electrical systems. The development of new elements such as the inerter or the memristor is a clear example.
However, new applications and possibilities of using these analogues still remain to be explored.
In this work, the electrical analogues of different vehicle models are presented. A new and not previously reported analogy  between inertial coupling and electrostatic capacitive coupling is found and described. Several examples are provided to highlight the benefits of this analogy. Well-known mechanical systems like the half-car or three three-axle vehicle models are discussed and some numerical results are presented. To the best of the author's knowledge, such systems were never dealt with by using a full electromechanical analogy. The mechanical equations are also derived and compared with those of the electrical domain for harmonic steady state analysis.  

\end{abstract}

\begin{keyword}
electromechanical analogy, capacitive coupling, vertical dynamics, vehicle model.
\end{keyword}

\end{frontmatter}

\section{Introduction}

Analogies can be established between different physical domains, such as mechanical, electrical, fluid or thermal systems, since they are modeled with comparable differential equations. Dynamical analogies are based on energy relations. In \cite{jeltsema2009multidomain}, Jeltsema \& Scherpen provide an overview of both the energy- and power-based modeling frameworks in different physical domains and discuss their mutual relationships.

Mechanical-electrical analogy was developed and rather extensively used in 30-40’s for the study of vibrations in linear mechanical systems \cite{firestone1933new, olson1943dynamical, bloch1945electromechanical, Mason1942, Miles1943}, where the use of mechanical to electrical analogy was empowered by the existing solution methods for electrical networks.
One of the earlier works that solved a mechanical system using electrical network theory was written by Harrison \cite{harrison1929electromagnetic} in a patent of invention published in 1929.
A detailed overview  with interesting historical notes of the conception and evolution of electromechanical analogy can be found in the work of Gardonio \& Brennan \cite{gardonio2002origins}.

Two different analogues have been used to translate mechanical systems into  electrical ones. Historically, the first proposed analogy  related force to voltage. In this so-called “force-voltage” analogy \cite{olson1943dynamical} (also known as “direct analogy”), the mechanical mass is related with an electrical inductor and the mechanical spring with an electrical capacitor. Few years after this analogy had been adopted, some difficulties or limitations were pointed out in \cite{firestone1933new} due to the fact that the physical interpretation of the electric network analogue was not direct from the mechanical system. In the force-voltage analogy, the relationship between mechanical and electrical elements do not preserve the same topology, i.e. mechanical elements arranged in series (parallel) are represented by electrical elements arranged in parallel (series). Moreover, the concept of \textit{through} and \textit{across} variables are inverted. A \textit{through} variable is measured on a single point of an element, as performed with force and current measures, while the value of an \textit{across} variable is obtained as the difference between the measurements in two different points, such is the case of velocity and voltage \cite{Miles1943}. Then, in the  force-voltage analogy, a force being a \textit{through} variable in the mechanical system  corresponds to a voltage, which is an \textit{across} variable in the electrical system.

To overcome these constraints, the new "force-current" analogy was formulated \cite{firestone1933new,hahnle1932darstellung}, and  the concept of \textit{bar impedance} (later on renamed as \textit{mobility}) was introduced \cite{firestone1938mobility}. This alternative analogy, also known as “inverse analogy”, preserves the same topology for both mechanical and  electrical systems, while keeping the equivalence between \textit{through} and \textit{across} variables.

In spite of the above, any of the described analogies are mathematically valid and may be applied indistinctly. Furthermore, depending on the specific mechanical problem, one analogy may be more appropriate over the other and easier to derive \cite{bloch1945electromechanical,gardonio2002origins}.
In other cases, it could be even  necessary to use both analogies to draw different parts of a mechanical system. The different electric diagrams derived are then linked using appropriated couplers \cite{bauer1953transformer}.
It should be noted that the two possible electric analogues obtained for a given mechanical system (force-voltage and force-current) are dual to each other. Then, it is possible to transform one into another following some basic relations \cite{bloch1945electromechanical}. As in the electric network, the duality principle holds for mechanical systems \cite{le1952duality}. Some illustrative examples of dual mechanical system  can be found in \cite{gardonio2000mobility}.

Beyond purely mechanical systems, the utility of analogues is especially remarkable when mechanical systems are linked to electrical systems. In this multidomain problem, the mechanical system is replaced by its electrical equivalent and joined to the electrical one. In this way and unique electrical system is studied \cite{falaize2020passive, tiwari2017lumped}.
De Silva~\cite{deSilva2010use} proposed the use of linear graphs for modeling multi-domain systems in a unified way, thus allowing to exploit the existing analogies across domains. More recently, de~Silva~\cite{deSilva2017} introduced a systematic approach for modeling multi-domain systems in a “unique” (single) model having physically meaningful variables, and many illustrative examples were described.

Mechanical systems can be directly drawn as mechanical “circuits” or diagrams using mechanical symbols, instead of using an electrical representation, but where the electric principles can be applied. This approach is known as the “mobility method” when the same topology of the mechanical system is preserved (as in the force-current analogy), or “impedance method” when the topology is changed as in the force-voltage analogy \cite{firestone1938mobility,gardonio2000mobility}. Firestone \cite{firestone1938mobility} stated the equivalent of the Kirchhoff’s laws for the mobility method: (i) Force Law: the sum of all the forces acting on any junction point is zero; (ii) Velocity Law: the sum of all the velocities across the structures included in any closed mechanical circuit is zero. The solution of this mechanical diagrams is then obtained without any reference to electric systems \cite{plunkett1958colloquium}.

Working with analogues facilitates the transfer of knowledge and ideas between the different branches of science and engineering. It motivates scientists to become interested in other fields, to create synergies and interdisciplinary working groups. Also, in the educational field, the use of analogues helps at approaching and understanding of the different subjects \cite{deSilva2019systematic}.
Analogues allow solving problems of one physical system by using resolution methods of another physical system that may show some advantage. In this sense, it has been more often preferred to work with the electrical analogues, while there are interesting works where the mechanical analogue of electrical power system was used, see references ~\cite{Bergvall1928915}, \cite{Reich1932287} and \cite{Pawley1937179}.

Analogies between different physical domains have helped at finding new elements. This is the case of the memristor. In the electrical domain, a memristor is a two terminal circuit element characterized by a relationship between the charge and the flux linkage. The existence of that missing constitutive relation was described by Chua in 1971 \cite{chua1971memristor}, though it was not until 2008 when an electrical passive memristive device was constructed \cite{strukov2008missing}.  One year after the work of Chua, Oster \& Auslander \cite{oster1972memristor} proposed a tapered dashpot as a mechanical memristor, showing a relation between displacement and momentum, which are the mechanical analogues of electric charge and flux linkage.
Another remarkable contribution attributable to the use of analogues is the invention of the inerter \cite{smith2002synthesis}. The inerter was the result of searching for a genuine two-terminal mechanical device equivalent to the electrical capacitor 
Unlike a conventional mass element, the electrical equivalent of the inerter does not require a grounded terminal. The inerter is the true dual of the spring.

Eletro-mechanical analogies have been used in vehicle suspension modeling and control \cite{pehlivan2018modelingm,xu2013modeling,jiamei2010suspension,shen2019optimal}, in vehicle drive trains \cite{routex2002study}, in structural dynamics \cite{fahy2018advanced}, in  modeling and control of flexible structures \cite{lenning1997integration}, in  design and optimization of inductive power transfer systems \cite{luca2017inductive}, and in piezoelectric vibration energy harvesters \cite{liao2019unified}, among others mechanical or electromechanical systems with interest in vibrations \cite{falaize2020passive, tiwari2017lumped, lossouarn2016multimodal, bai2003design, mamedov2010mathematical}.

\section{Contributions}
The present work is a contribution to the use of electromechanical analogues in vehicle suspensions. To the authors’ knowledge, the use of electric analogues   in the scientific literature regarding vehicle suspensions is limited to the quarter-car model. In this work, the full electromechanical analogue of a half-car vehicle model,  where inertial coupling appears due to the vehicle main frame, is presented. The relationship between mechanical inertial coupling and electrostatic capacitor coupling is then described. This is a novelty of this work in the quest of identifying new analogies. Furthermore, the analogy is extended to a three-axle vehicle model in which some numerical results are obtained and discussed.



\section{Electrical analogies of  mechanical systems. A basic example} 
\begin{figure}
    \centering
    \subfigure []{ 
\begin{tikzpicture}[every node/.style={outer sep=0pt},thick,
    		mass/.style={draw,thick},
    		spring/.style={thick,decorate,decoration={zigzag,pre length=0.3cm,post length=0.3cm,segment length=6}},
    		ground/.style={fill,pattern=north east lines,draw=none,minimum width=0.75cm,minimum height=0.3cm},
    		dampic/.pic={\fill[white] (-0.1,-0.3) rectangle (0.3,0.3);
    				\draw (-0.3,0.3) -| (0.3,-0.3) -- (-0.3,-0.3);
    				\draw[line width=1mm] (-0.1,-0.3) -- (-0.1,0.3);},
				masspic/.pic={\fill[white] (-0.4,-0.3) rectangle (0.3,0.3);
										\draw (-0.4,0.3) |- (0.3,-0.4);
										\draw (-0.3,-0.3) -| (0.3,0.3) -| (-0.3,-0.3);}]
    
      \node[mass,minimum width=1.0cm,minimum height=1.25cm,fill=white!80!black] (m1) {$m_1$};
    	\node[right=1.5cm of m1,mass,minimum width=1.0cm,minimum height=1.25cm,fill=white!80!black] (m2) {$m_2$};
      \node[left=1.5cm of m1,ground,minimum width=3mm,minimum height=2.5cm] (g1){};
      \draw (g1.north east) -- (g1.south east);
    
      \draw[spring] ([yshift=4mm]g1.east) coordinate(aux) -- (m1.west|-aux) node[midway,above=1mm]{$k_1$};
      \draw[spring] ([yshift=4mm]m1.east) coordinate(aux) -- (m2.west|-aux) node[midway,above=1mm]{$k_2$};

      \draw ([yshift=-4mm]g1.east) coordinate(aux') -- (m1.west|-aux') pic[midway]{dampic} node[midway,above=2.5mm]{$d_1$}
    				([yshift=-4mm]m1.east) coordinate(aux') -- (m2.west|-aux') pic[midway]{dampic} node[midway,above=2.5mm]{$d_2$};

      \foreach \X in {1,2}  
      {\draw[thin] (m\X.south) -- ++ (0,-1) coordinate[pos=0.85](aux'\X);
       \draw[latex-] (aux'\X) -- ++ (-1,0) node[midway,above]{$x_\X$}
        node[left,ground,minimum height=7mm,minimum width=1mm] (g'\X){};
       \draw[thick] (g'\X.north east) -- (g'\X.south east);
      }
    	
    \draw[latex-] (m2.east) -- ++ (1,0) node[above left]{$f(t)$};
    
\end{tikzpicture}}\\
    \subfigure []{
\begin{tikzpicture}[every node/.style={outer sep=0pt},thick,
    		mass/.style={draw,thick},
    		spring/.style={thick,decorate,decoration={zigzag,pre length=0.3cm,post length=0.3cm,segment length=6}},
    		ground/.style={fill,pattern=north east lines,draw=none,minimum width=0.75cm,minimum height=0.3cm},
    		dampic/.pic={\fill[white] (-0.1,-0.3) rectangle (0.3,0.3);
    				\draw (-0.3,0.3) -| (0.3,-0.3) -- (-0.3,-0.3);
    				\draw[line width=1mm] (-0.1,-0.3) -- (-0.1,0.3);},
				masspic/.pic={\fill[white] (-0.4,-0.3) rectangle (0.3,0.3);
										\draw (-0.4,0.3) |- (0.3,-0.4);
										\draw (-0.3,-0.3) -| (0.3,0.3) -| (-0.3,-0.3);}]

      \draw[thick] (0,-1) -- (0,1) (m1) {};
      \draw[thick] (2.5,-1) -- (2.5,1) (m2) {};
      \node[left=1.5cm of m1,ground,minimum width=3mm,minimum height=2.5cm] (g1){};
      \draw (g1.north east) -- (g1.south east);
    
      \draw[spring] ([yshift=8mm]g1.east) -- (0,8mm) node[midway,above=1mm]{$k_1$};
      \draw[spring] (0,8mm) -- (2.5,8mm) node[midway,above=1mm]{$k_2$};

      \draw ([yshift=0mm]g1.east) -- (0,0) pic[midway]{dampic} node[midway,above=2.5mm]{$d_1$}
    				(0,0) -- (2.5,0) pic[midway]{dampic} node[midway,above=2.5mm]{$d_2$};
						
			\draw ([yshift=-8mm]g1.east) coordinate(aux') -- (0,-8mm) pic[midway]{masspic} node[midway]{$m_1$}
    				(0.6,-8mm) -- (2.5,-8mm) pic[midway]{masspic} node[midway]{$m_2$};
         
    \pattern[pattern=north east lines] (0.3,-12mm) rectangle (0.6,-4mm);
    \draw[thick] (0.6,-12mm) -- (0.6,-4mm);

    \draw (3.75,0) to[thin,american current source, label=\mbox{$f(t)$}] (2.5,0);
    \pattern[pattern=north east lines] (3.75,-0.25) rectangle (4.0,0.25);
    \draw[thick] (3.75,-0.25) -- (3.75,0.25);		

    \end{tikzpicture}
 }
    \caption{(a) 2 d.o.f. mechanical model and (b) its network representation.}
    \label{fig:model2gdl}
\end{figure}
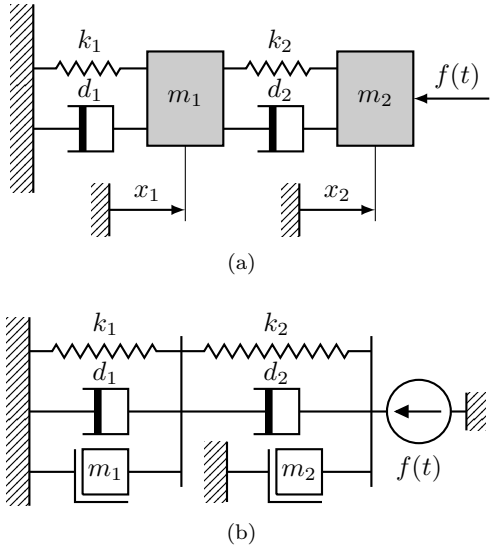

The process followed to obtain an electric analogy of a mechanical system is shown in this section with the help of the 2 d.o.f translational model depicted in figure~\ref{fig:model2gdl}. Masses $m_1$ and $m_2$ are linked by a spring and a damper arranged in parallel, with $k_2$ and $d_2$ as stiffness and damping constants, respectively. Mass $m_1$ is connected to the ground by another parallel spring-damper pair with stiffness and damping constants $k_1$ and $d_1$, respectively. An external force $f(t)$ is acting on the mass $m_2$. Using the mobility method \cite{gardonio2000mobility}, figure~\ref{fig:model2gdl}b shows the mechanical network of the model, where masses $m_1$ and $m_2$ are “connected" to ground due to the inertial frame of reference, i.e., the velocity and acceleration of these masses are measured relative to ground.

The equations of motion of the system in figure~\ref{fig:model2gdl}a can be obtained by using Lagrange equations~\cite{gantmakher1970lectures} in terms of the independent variables $x_1$ and $x_2$, which are the components of the state vector $\vc{x}=\left[x_1\;\;x_2\right]^T$ as follows:
\begin{equation}
\dfrac{\rm d}{\mathrm{d}t}\left(\dfrac{\partial L}{\partial \vc{\dot{x}}}\right)-\dfrac{\partial L}{\partial \vc{x}}+
\dfrac{\partial F_R}{\partial \vc{\dot{x}}} = \vc{Q}_a(t)
\label{eq:lagrange_ecus}
\end{equation}
where $L=T-\Pi$ is the lagrangian function, $T$ is the kinetic energy, $\Pi$ is the potential energy, $F_R$ is a Rayleigh dissipation function and $\vc{Q}_a(t)$ is the generalized applied force vector. The expressions of $T$, $\Pi$ and $F_R$ read as follows:
\begin{align}
&T = \frac{1}{2}m_1\dot{x}_1^2 + \frac{1}{2}m_2\dot{x}_2^2\\
&\Pi = \frac{1}{2}k_1 x_1^2 + \frac{1}{2}k_2 \left(x_2-x_1\right)^2\\
&F_R = \frac{1}{2}d_1\dot{x}_1^2 + \frac{1}{2}d_2\left(\dot{x}_2-\dot{x}_1\right)^2
\end{align}
Introducing $T$, $\Pi$, $F_R$ and $\vc{Q}_a(t) = \left[0\;\; -f(t)\right]^T$ into Equation~\eqref{eq:lagrange_ecus}, the following ODE system is found:
\begin{equation}\begin{split}
&\underbrace{\matr{m_1}{0}{0}{m_2}}_{\vc{m}}\underbrace{\vect{\ddot{x}_1}{\ddot{x}_2}}_{\vc{\ddot{x}}} +
\underbrace{\matr{d_1+d_2}{-d_2}{-d_2}{d_2}}_{\vc{d}}\underbrace{\vect{\dot{x}_1}{\dot{x}_2}}_{\vc{\dot{x}}} +\\
&\underbrace{\matr{k_1+k_2}{-k_2}{-k_2}{k_2}}_{\vc{k}}\underbrace{\vect{x_1}{x_2}}_{\vc{x}} =
\underbrace{\vect{0}{-f(t)}}_{\vc{Q}(t)}
\end{split}
\label{eq:2gdl_time_mec_eq}
\end{equation}
which may be rewritten in matrix form as
\begin{equation}
\vc{m}\vc{\ddot{x}} + \vc{d}\vc{\dot{x}} + \vc{k}\vc{x} = \vc{Q}(t) \label{eq:equ_structure}
\end{equation}
%
Note that  $\vc{Q}(t)=\vc{Q}_a(t)$ in \eqref{eq:equ_structure}, but this is not always the case, as it will be shown for vehicle model subject to excitations coming from the road profile.

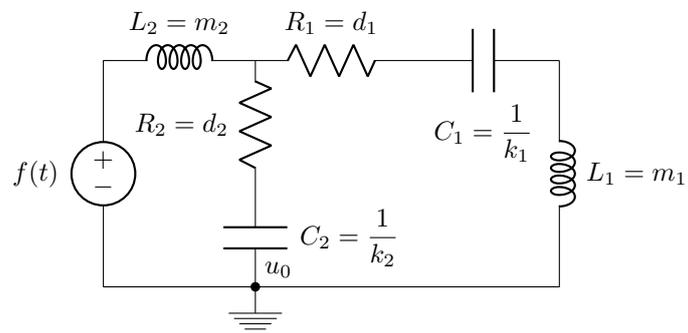
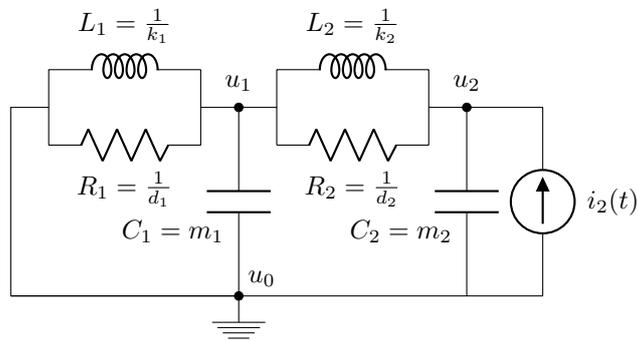
\begin{figure}
	\centering
	\subfigure[]{ 
	\begin{circuitikz}[scale=2]
    \def\xa{0}
    \def\xb{1}
		\def\xc{2}
		\def\xd{3}

		\def\ya{0}
    \def\yb{0.65}
    \def\yc{1.50}

    \draw           (\xa,\ya)
	        to[american voltage source, label=\mbox{$f(t)$}, invert] (\xa,\yc)	
					to[inductor,l = ${L_2=m_{2}}$]   (\xb,\yc);
		\draw   (\xb,\ya)  to[capacitor,l_= ${C_2=\dfrac{1}{k_2}}$] (\xb,\yb) to[resistor, l=${R_2 = d_2}$] (\xb,\yc);	
		\draw   (\xb,\yc)  to[resistor, l=${R_1=d_1}$] (\xc,\yc) to[capacitor,l_= ${C_1=\dfrac{1}{k_{1}}}$] (\xd,\yc);  	
		\draw   (\xd,\yc) to[inductor,l= ${L_1=m_1}$] (\xd,\ya) to[short] (\xa,\ya);
\draw (\xa,\ya) to [short,-*] (\xb,\ya) node[ground, label={60:$u_0$}]{};

\end{circuitikz} 
	}
	\hspace{0mm}
	\subfigure[]{ 
\begin{circuitikz}[scale=2]
    \def\xa{0}
    \def\xb{0.25}
		\def\xc{1.25}
		\def\xd{1.5}
		\def\xe{1.75}
		\def\xf{2.75}
		\def\xg{3}
		\def\xh{3.5}

		\def\ya{0}
    \def\yb{1.0}
    \def\yc{1.25}
    \def\yd{1.50}

    \draw                 (\xa,\ya)
	        to[short] 			(\xa,\yc)	
					to[short]				(\xb,\yc);
    \draw                 (\xc,\yc) 
            to[short] 									(\xe,\yc)
						to[short]										(\xe,\yd)
						to[inductor,l = ${L_{2}=\frac{1}{k_2}}$]  (\xf,\yd)
						to[short]										(\xf,\yb)
						to[resistor,l= ${R_{2}=\frac{1}{d_2}}$]   (\xe,\yb)
						to[short]										(\xe,\yc);
		\draw    														 (\xb,\yc)
            to[short]                    (\xb,\yd)
            to[inductor,l = ${L_{1}=\frac{1}{k_1}}$]   (\xc,\yd)
            to[short]                    (\xc,\yb)
            to[resistor,l= ${R_{1}=\frac{1}{d_1}}$]    (\xb,\yb)
            to[short]                    (\xb,\yc);

\draw   (\xd,\ya)  to[C,-*] (\xd,\yc) node[above=1mm]{{$u_1$}};
\node[below left=0.1cm,font=\normalsize] at (\xd,0.6) {${C_{1}=m_1}$};		
\draw   (\xg,\ya) to[C,-*] (\xg,\yc) node[above=1mm]{{$u_2$}};
\node[below left=0.1cm,font=\normalsize] at (\xg,0.6) {$C_2=m_2$};

\draw (\xf,\yc) to (\xh,\yc)
          to[american current source, label=\mbox{$i_2(t)$}, invert] (\xh,\ya)
          to (\xa,\ya);

\draw (\xa,\ya) to [short,-*] (\xd,\ya) node[ground, label={60:$u_0$}]{};

\end{circuitikz}

	} 
	\caption{(a) Force-voltage and (b) force-current analogues of the 2 d.o.f. mechanical model of figure \ref{fig:model2gdl}.}
	\label{fig:analog2gdl}
\end{figure}

This mechanical system can be studied through any of the two well-known versions of electromechanical analogies. Table \ref{tab:tabla1} summarizes the relationship between the electrical and mechanical elements and variables for the force-voltage and force-current analogies. Both electrical analogues are shown in figure~\ref{fig:analog2gdl}, where some basic connection rules have been followed \cite{bloch1945electromechanical}. Mainly, when force-voltage analogy is used, parallel (series) connections in the mechanical systems must be drawn as series (parallel) connections in the electrical network. Conversely,  if force-current analogy is preferred, the topology of the diagrams is not altered \cite{Miles1943}. The later is a very strong argument in favour of the force-current analogy. Furthermore, it will also facilitates the resolution of the proposed networks.  
Therefore, the force-current version will be used in the rest of this work. 

Figure \ref{fig:analog2gdl}b represents the force-current analog of the mechanical system of figure \ref{fig:model2gdl}. It can be seen that the topology of the electrical circuit is identical to the mechanical one. It is worth to mention that capacitors must always have a lead to the common node or earth. As explained before, a mass that moves respect to the ground (the inertial frame), behaves like a capacitor tied to  ground in the electrical network. 

\begin{table}[]
\centering
\begin{tabular}{lllrlr}
\toprule
  &   & \multicolumn{4}{c}{\textbf{Electrical system analogy}}      \\\cmidrule{3-6}
\multicolumn{2}{c}{\textbf{Mechanical variables}}     & \multicolumn{2}{c}{\textbf{Force-voltage}}       & \multicolumn{2}{c}{\textbf{Force-current}}     \\\cmidrule(lr){1-2}\cmidrule(lr){3-4}\cmidrule(lr){5-6}
Force [N]      & $F$ &  $u$     & Voltage [V]     &  $i$    & Current [A]      \\
Velocity [m/s] & $v$ &  $i$    & Current [A]      &   $u$    & Voltage [V]     \\
Spring [N/m]   & $k$ &  $C=1/k$   & Capacitor [F] &  $L=1/k$   & Inductor [H]    \\
Mass (inerter) [kg]      & $m$ &   $L=m$   & Inductor [H]   &  $C=m$  & Capacitor [F]  \\
Damper [Ns/m] & $d$ & $R=d$ & Resistor [$\Omega$] & $R=1/d$ & Resistor [$\Omega$] \\ \bottomrule
\end{tabular}%
\caption{General analogy between mechanical and electrical systems.}
\label{tab:tabla1}
\end{table}

Note that once the topology of the electrical analogue is found, it can be used for either time  or frequency domain analysis. Based on table \ref{tab:tabla1} and the use of Kirchhoff's current law (KCL), the following equation can be obtained:

\begin{equation}
\vc{C}\vc{\ddot{\phi}} + \vc{G}\vc{\dot{\phi}} + \vc{B}\vc{\phi} = \vc{I}(t) \label{eq:equ_structure_elec}
\end{equation}

which can be expanded to:

\begin{equation}\begin{split}
&\matr{C_1}{0}{0}{C_2}\vect{\ddot{\phi}_1}{\ddot{\phi}_2} +
\matr{G_1+G_2}{-G_2}{-G_2}{G_2}\vect{\dot{\phi}_1}{\dot{\phi}_2} +
\matr{B_1+B_2}{-B_2}{-B_2}{B_2}\vect{\phi_1}{\phi_2} =
\vect{0}{-i_2(t)}
\end{split}
\label{eq:2gdl_time_elec_eq}
\end{equation}

\noindent where 
$G_i=\frac{1}{R_i}$ is the conductance of resistor $R_i$ and $B_i=\frac{1}{L_i}$ is the inverse of inductance $L_i$. The variable $\phi$ is the flux linkage. Note that $\dot{\phi}_j=u_j$ is the voltage of node $j$ and the force $f(t)$ has been replaced by its analogue $i(t)$. It can be readily observed that (\ref{eq:equ_structure_elec}) and (\ref{eq:2gdl_time_elec_eq}) are similar to (\ref{eq:equ_structure}) and (\ref{eq:2gdl_time_mec_eq}), respectively. For steady state harmonic analysis, Equation (\ref{eq:2gdl_time_elec_eq}) can be transferred to the frequency domain by using the Euler expression $e(t)=Re\{\sqrt{2}\vec{E}e^{j\omega t}\}$, where $\vec{E}=E e^{j\varphi}$ is widely-known as a phasor. It indicates the RMS ($E$) and phase ($\varphi$) value of the harmonic sinusoid, respectively. Thus, Equation (\ref{eq:2gdl_time_elec_eq}) can be written in a compact way as follows:

\begin{gather}
    \begin{bmatrix}
     0 \\ \vec{I}_2
    \end{bmatrix}
    =
       \underbrace{\begin{bmatrix}
      G_1+\vec{B}_{L_1}+\vec{B}_{C_1}+G_2+\vec{B}_{L_2} & -G_2-\vec{B}_{L_2}\\
      -G_2-\vec{B}_{L_2} & \vec{B}_{C_2}+G_2+\vec{B}_{L_2}
    \end{bmatrix}}_{\vec{Y}}
       \underbrace{\begin{bmatrix}
     \vec{U}_1 \\ \vec{U}_2
    \end{bmatrix}}_{\vec{U}}
    \label{eq:sistema_mob_nudos}
\end{gather}

\noindent where  $\vec{B}_{L_i}=\frac{1}{L_i\omega j}$ is the susceptance (inverse of complex reactance) associated to $L_i$ and $\vec{B}_{C_i}=C_i\omega j$ is the susceptance associated to $C_i$.
The matrix system in (\ref{eq:sistema_mob_nudos}) is composed of complex numbers and can be easily solved by using matrix linear algebra. The unknown voltages $\vec{U}_1$ and $\vec{U}_2$ are obtained by inverting the admittance matrix $[\vec{Y}]$:

\begin{equation}
    [\vec{U}] = [\vec{Y}]^{-1} [\vec{I}]
\end{equation}

Once the voltages are known, every current can be solved by applying Ohm's law to any element of the circuit. This is equivalent to know the velocity and force in every element of the original mechanical system. The real advantage of the method resides in the application of the countless theorems and laws developed over the years in network analysis. For example, the dimensional reduction of the circuit can be realized by means of Thevenin/Norton theorem.

\section{Electromechanical analogue of a half-car model based on electrostatic capacitor coupling}
\begin{figure}%
\centering
\includegraphics[width=0.6\columnwidth]{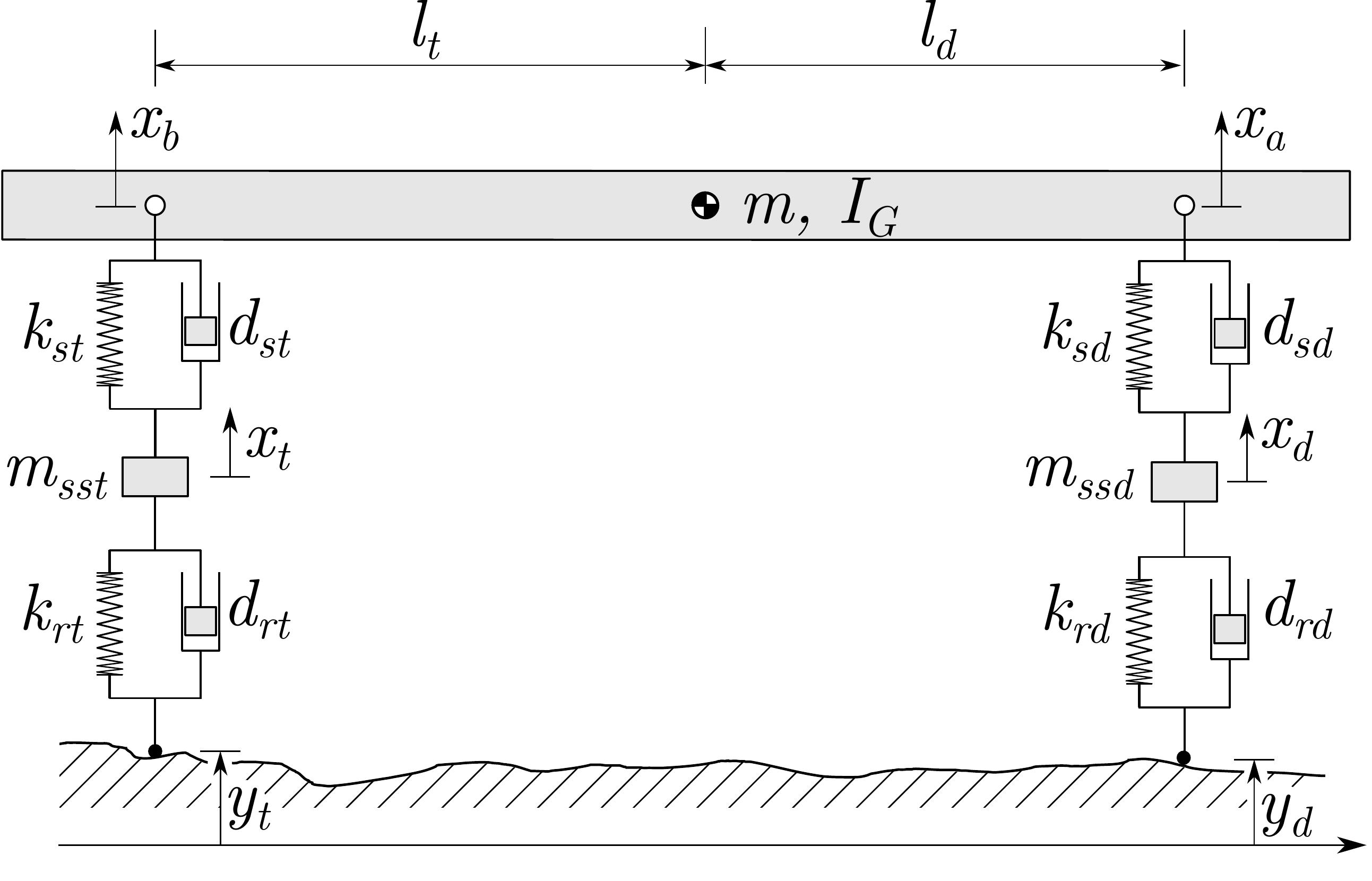}%
\caption{Half-car vehicle model for vertical dynamics analysis.}%
\label{fig:half_car}%
\end{figure}

In this section the analogy is applied to a more complex mechanical system like the classical half car vehicle model depicted in figure~\ref{fig:half_car}, showing elastic and inertial coupling. The equations of motion of such system could be written without inertial coupling terms if the vertical displacement of the center of mass and the pitch angle are taken as coordinates. 
Nevertheless, in order to have only translational coordinates, the vertical displacements of the main frame points where the suspensions are attached, $x_a$ and $x_b$, are used as coordinates. In this way, neither the mass nor the stiffness matrices are diagonal and, therefore, the system shows inertial and elastic coupling. 

The mechanical system in figure~\ref{fig:half_car} has four degrees of freedom. The coordinate vector is written as 
\begin{equation}
\vc{x}=\left[\begin{array}{cccc} {{x}_{a}} & {{x}_{b}} & {{x}_{d}} & {{x}_{t}} \end{array}\right]^T
\end{equation}
where ${{x}_{d}}$ and ${{x}_{t}}$ are the vertical displacements of the unsprung masses. The equations of motion will be obtained again by using Lagrange equation \eqref{eq:lagrange_ecus} for which we need to calculate the kinetic and potential energies as well as the Rayleigh dissipation function as follows: 
\begin{align}
T=& \frac{{m_{ssd}}{{\dot{x}_{d}}}^2}{2}+\frac{{m_{sst}}{{\dot{x}_{t}}}^2}{2}+\frac{{I_{G}}{\left({\dot{x}_a}-{\dot{x}_b}\right)}^2}{2{\left({l_{d}}+{l_{t}}\right)}^2}+\frac{m{\left({l_{d}}{\dot{x}_b}+{l_{t}}{\dot{x}_a}\right)}^2}{2{\left({l_{d}}+{l_{t}}\right)}^2}\\
\Pi =& \frac{{k_{sd}}{\left({{x}_{a}}-{{x}_{d}}\right)}^2}{2}+\frac{{k_{st}}{\left({{x}_{b}}-{{x}_{t}}\right)}^2}{2}+\frac{{k_{rd}}{\left({{x}_{d}}-{{y}_{d}}\right)}^2}{2}+\nonumber\\&\frac{{k_{rt}}{\left({{x}_{t}}-{{y}_{t}}\right)}^2}{2}+g{m_{ssd}}{{x}_{d}}+g{m_{sst}}{{x}_{t}}+\frac{gm\left({l_{d}}{{x}_{b}}+{l_{t}}{{x}_{a}}\right)}{{l_{d}}+{l_{t}}}\\
F_R =& \dfrac{{d_{sd}}\,{\left({\dot{x}_a}-{\dot{x}_{d}}\right)}^2}{2}+\frac{{d_{st}}{\left({\dot{x}_b}-{\dot{x}_{t}}\right)}^2}{2}+\frac{{d_{rd}}{\left({\dot{x}_{d}}-{\dot{y}_{d}}\right)}^2}{2}+\nonumber\frac{{d_{rt}}\,{\left({\dot{x}_{t}}-{\dot{y}_{t}}\right)}^2}{2}
\end{align}
where $m$ and $I_G$ are the mass and moment of inertia of the vehicle frame, $m_{ssd}$ and $m_{sst}$ are the front and rear unsprung masses, $k_{sd}$, $d_{sd}$, $k_{st}$, $d_{st}$ are the stiffness and damping constants of the front and rear suspension elements, $k_{rd}$, $d_{rd}$, $k_{rt}$, $d_{rt}$ are the front and rear tyre stiffness and damping constants and ${{y}_{d}}$ and ${{y}_{t}}$ are the front and rear displacements of the wheel and ground contact points. Resorting to Equation~\eqref{eq:equ_structure}, the mass, damping and stiffness matrices read as follows:
\begin{equation}
\vc{m}=\left[\begin{array}{cccc} \dfrac{m\,{{l_{t}}}^2+{I_{G}}}{{\left({l_{d}}+{l_{t}}\right)}^2} & \dfrac{{l_{d}}\,{l_{t}}\,m-{I_{G}}}{{\left({l_{d}}+{l_{t}}\right)}^2} & 0 & 0\\ \dfrac{{l_{d}}\,{l_{t}}\,m-{I_{G}}}{{\left({l_{d}}+{l_{t}}\right)}^2} & \dfrac{m\,{{l_{d}}}^2+{I_{G}}}{{\left({l_{d}}+{l_{t}}\right)}^2} & 0 & 0\\ 0 & 0 & {m_{ssd}} & 0\\ 0 & 0 & 0 & {m_{sst}} \end{array}\right],
\label{eq:m_matrix_halfcar_coupling}
\end{equation}

\begin{equation}
\vc{d}=\left[\begin{array}{cccc} {d_{sd}} & 0 & -{d_{sd}} & 0\\ 0 & {d_{st}} & 0 & -{d_{st}}\\ -{d_{sd}} & 0 & {d_{rd}}+{d_{sd}} & 0\\ 0 & -{d_{st}} & 0 & {d_{rt}}+{d_{st}} \end{array}\right],
\end{equation}

\begin{equation}
\vc{k}=\left[\begin{array}{cccc} {k_{sd}} & 0 & -{k_{sd}} & 0\\ 0 & {k_{st}} & 0 & -{k_{st}}\\ -{k_{sd}} & 0 & {k_{rd}}+{k_{sd}} & 0\\ 0 & -{k_{st}} & 0 & {k_{rt}}+{k_{st}} \end{array}\right],
\end{equation}
and the generalized force vector is written as follows
\begin{equation}
\vc{Q}(t)=\left[\begin{array}{c} -\dfrac{g\,{l_{t}}\,m}{{l_{d}}+{l_{t}}}\\ -\dfrac{g\,{l_{d}}\,m}{{l_{d}}+{l_{t}}}\\ -g\,{m_{ssd}}\\ -g\,{m_{sst}} \end{array}\right] + \left[\begin{array}{c} 0\\ 0\\ {d_{rd}}\,{\dot{y}_{d}}(t)+{k_{rd}}\,{{y}_{d}}(t)\\ {d_{rt}}\,{\dot{y}_{t}}(t)+{k_{rt}}\,{{y}_{t}}(t) \end{array}\right]
\label{eq:equations_Q_half_car}
\end{equation}
where $\vc{Q}(t)$ is the sum of a gravitational force vector and a non-constant vector dependent on the road profile. For harmonic steady state analysis, gravitational forces can be omitted as they result in an offset that may be added if needed.
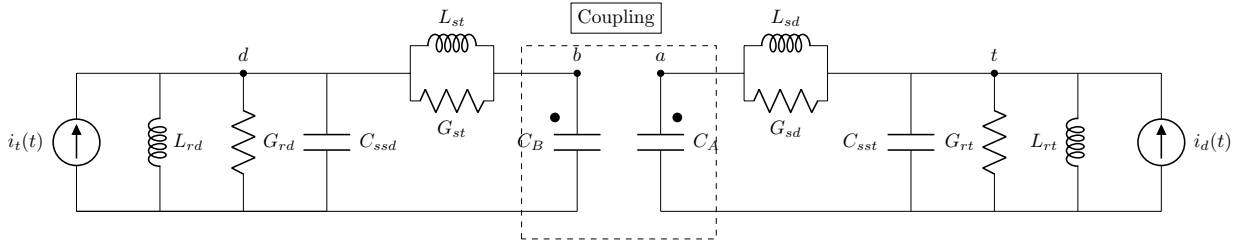
\begin{figure*}%
	\resizebox{\linewidth}{!}{
\begin{circuitikz}[scale=2]
    \def\xa{0}
    \def\xb{0.75}
		\def\xc{1.5}
		\def\xd{2.25}
		\def\xe{3}
		\def\xf{3.75}
		\def\xg{4.5}
		\def\xh{5.25}
		\def\xi{6}
		\def\xj{6.75}
		\def\xk{7.5}
		\def\xl{8.25}
		\def\xm{9}
		\def\xn{9.75}

		\def\ya{0}
    \def\yb{1.00}
    \def\yc{1.25}
    \def\yd{1.50}

    \draw                               (\xa,\ya)
						to[american current source, label=\mbox{$i_t(t)$}] (\xa,\yc)
						to[short]										(\xe,\yc)
						to[short]										(\xe,\yd)
						to[inductor,l = {$L_{st}$}]   (\xf,\yd)
						to[short]										(\xf,\yb)
						to[resistor,l= {$G_{st}$}]    (\xe,\yb)
						to[short]										(\xe,\yc);
		\draw                               (\xf,\yc)			
						to[short]										(\xg,\yc);
		\draw			(\xh,\yc)	to[short]										(\xi,\yc)
						to[short]										(\xi,\yd)
						to[inductor,l = {$L_{sd}$}]   (\xj,\yd)
						to[short]										(\xj,\yb)
						to[resistor,l= {$G_{sd}$}]    (\xi,\yb)
						to[short]										(\xi,\yc);
		\draw                               (\xj,\yc)			
						to[short]										(\xn,\yc)
						to[american current source, label=\mbox{$i_d(t)$},invert] (\xn,\ya)
						to[short]										(\xh,\ya);
        \draw           to[short] (\xg,\ya);
							
		\draw   (\xb,\ya) to[inductor,l_= {$L_{rd}$}] (\xb,\yc);
		\draw   (\xc,\ya) to[resistor,l_= {$G_{rd}$},-*] (\xc,\yc) node[above=1mm]{{$d$}};
		\draw   (\xd,\ya) to[capacitor,l_= {$C_{ssd}$}] (\xd,\yc);
		
		\draw   (\xg,\ya) to[capacitor,l= {$C_{B}$},-*] (\xg,\yc) node[above=1mm]{{$b$}};
		\draw   (\xh,\ya) to[capacitor,l_= {$C_{A}$},-*] (\xh,\yc) node[above=1mm]{{$a$}};
		
		\draw   (\xk,\ya) to[capacitor,l= {$C_{sst}$}] (\xk,\yc);				
		\draw   (\xl,\ya) to[resistor,l= {$G_{rt}$},-*] (\xl,\yc) node[above=1mm]{{$t$}};
		\draw   (\xm,\ya) to[inductor,l= {$L_{rt}$}] (\xm,\yc);
		\draw   (\xa,\ya) to[short] (\xd,\ya);
		\draw[dashed] (4,-0.25) -- (4,1.5) -- (5.75,1.5) -- (5.75,-0.25) -- (4,-0.25);
        \node[draw] at (4.85,1.75) {Coupling};
  \node[circle,fill=black,inner sep=0pt,minimum size=5pt] (a) at (4.3,0.85) {};
  \node[circle,fill=black,inner sep=0pt,minimum size=5pt] (a) at (5.4,0.85) {};
\end{circuitikz}
}
\caption{Electrical analogue of the half-car vehicle model in figure~\ref{fig:half_car}. The vertical velocity of each wheel is modelled through voltages sources. Both the inertial and elastic coupling can be modelled through an electrostatic capacitor coupling.}%
\label{fig:half_car_coupling}%
\end{figure*}

By inspecting Equation (\ref{eq:m_matrix_halfcar_coupling}), it can be noticed that non-zero terms appear outside the diagonal. Based on circuit analysis techniques, it follows that there is an electrical coupling between variables $\ddot{\phi}_1$ and $\ddot{\phi}_2$, i.e., $\dot{u}_1$ and $\dot{u}_2$. Since we are using a force-current analogy, the right electrical analogue is depicted in figure \ref{fig:half_car_coupling}. Note that the same naming convention has been retained for nodes.  It can be observed that this analogue is very similar to that  in figure \ref{fig:analog2gdl}b  using the analogue twice, but introducing a new  mechanism based on electrostatic coupling between capacitors. Also, note that  the current source (the force $f(t)$) has been removed and  both, a new voltage source (to model the vertical velocity that causes the road profile) and an electrostatic capacitor coupling (to model the inertial and elastic coupling), have been included. The dots near the capacitors $C_A$ and $C_B$ in figure \ref{fig:half_car_coupling} indicates the polarity of the capacitor coupling. Interesting enough, this analogue is widely used in power electronics for different applications such as elecric vehicle charging \cite{yi2020capacitive}. Nota that the use of the force-current analogy leads to a dual of the classic magnetic coupling widely used for transformers and transducers.

\begin{figure}%
	\resizebox{\columnwidth}{!}{
\begin{circuitikz}[font=\sffamily, american currents]

\draw 
      (0,0) to [short] (3,0) to [cI] (3,3) to [short,-*] (1,3) to [short] (0,3);
\draw (1,0) to [capacitor, l={$C_B$}] (1,3);
\draw 
      (8,0) to [short] (5,0) to [cI] (5,3) to [short,-*] (7,3) to [short] (8,3);
\draw (7,0) to [capacitor, l={$C_A$}] (7,3);
\node[] at (4.3,0.9) {$j\omega C_M \vec{U}_b$};
\node[] at (3.9,2.1) {$j\omega C_M \vec{U}_a$};
\draw (1,3.25) node[above=1mm]{{$b$}};
\draw (7,3.25) node[above=1mm]{{$a$}};

\node at (9,1.5) {\Large $\equiv$};
\node at (9,1.9) {Equiv};

    \draw   (11,0) to[capacitor,-*] (11,3) node[above=1mm]{{$b$}};
    \draw   (14,0) to[capacitor,-*] (14,3) node[above=1mm]{{$a$}};

  \draw                               (10,3)      
            to[short]                    (11,3)
            to[capacitor,l= {$C_{M}$}]    (14,3)
            to[short]                    (15,3);
\draw (10,0) to[short] (15,0);
\node[] at (10.5,0.5) {$C_{B}-C_{M}=\dfrac{m l_d}{l}$};
\node[] at (13.8,2) {$C_{A}-C_{M}=\dfrac{m l_t}{l}$};

\end{circuitikz}

}
\caption{Equivalent networks for capacitive coupling. Left, voltage controlled current sources model and right, $\Pi$-network model.}%
\label{fig:coupling_models}%
\end{figure}
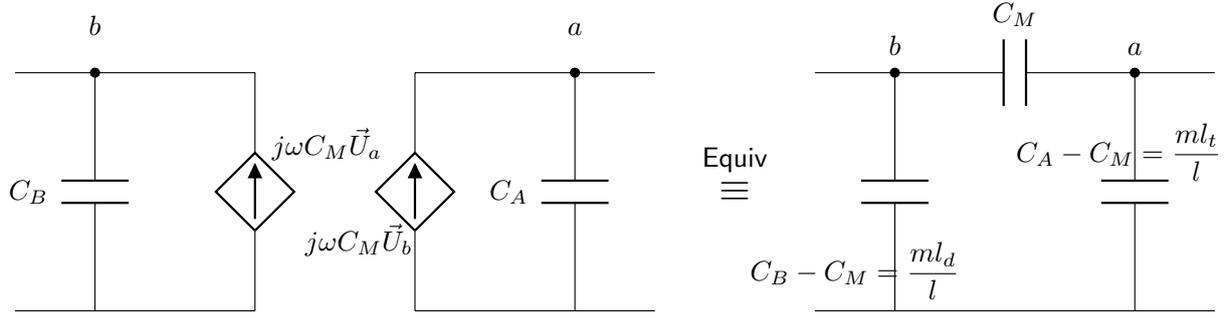

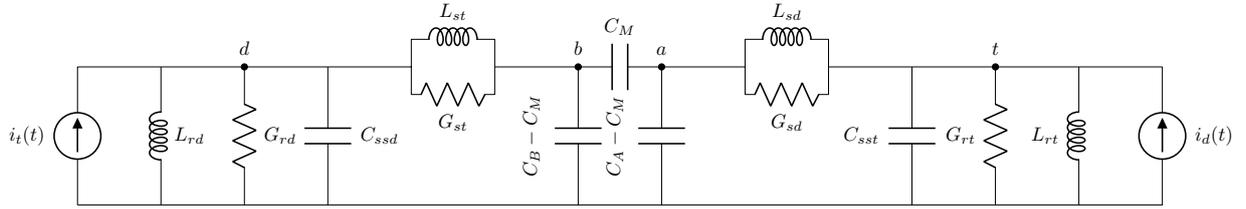
\begin{figure*}%
\resizebox{\textwidth}{!}{
\begin{circuitikz}[scale=2]
    \def\xa{0}
    \def\xb{0.75}
		\def\xc{1.5}
		\def\xd{2.25}
		\def\xe{3}
		\def\xf{3.75}
		\def\xg{4.5}
		\def\xh{5.25}
		\def\xi{6}
		\def\xj{6.75}
		\def\xk{7.5}
		\def\xl{8.25}
		\def\xm{9}
		\def\xn{9.75}

		\def\ya{0}
    \def\yb{1.00}
    \def\yc{1.25}
    \def\yd{1.50}

    \draw                               (\xa,\ya)
						to[american current source, label=\mbox{$i_t(t)$}] (\xa,\yc)
						to[short]										(\xe,\yc)
						to[short]										(\xe,\yd)
						to[inductor,l = {$L_{st}$}]   (\xf,\yd)
						to[short]										(\xf,\yb)
						to[resistor,l= {$G_{st}$}]    (\xe,\yb)
						to[short]										(\xe,\yc);
		\draw                               (\xf,\yc)			
						to[short]										(\xg,\yc)
						to[capacitor,l= {$C_{M}$}]		(\xh,\yc)
						to[short]										(\xi,\yc)
						to[short]										(\xi,\yd)
						to[inductor,l = {$L_{sd}$}]   (\xj,\yd)
						to[short]										(\xj,\yb)
						to[resistor,l= {$G_{sd}$}]    (\xi,\yb)
						to[short]										(\xi,\yc);
		\draw                               (\xj,\yc)			
						to[short]										(\xn,\yc)
						to[american current source, label=\mbox{$i_d(t)$},invert] (\xn,\ya)
						to[short]										(\xa,\ya);
							
		\draw   (\xb,\ya) to[inductor,l_= {$L_{rd}$}] (\xb,\yc);
		\draw   (\xc,\ya) to[resistor,l_= {$G_{rd}$},-*] (\xc,\yc) node[above=1mm]{{$d$}};
		\draw   (\xd,\ya) to[capacitor,l_= {$C_{ssd}$}] (\xd,\yc);
		
		\draw   (\xg,\ya) to[capacitor,label/align= rotate,l= {$C_{B}-C_{M}$},-*] (\xg,\yc) node[above=1mm]{{$b$}};
		\draw   (\xh,\ya) to[capacitor,label/align= rotate,l_= {$C_{A}-C_{M}$},-*] (\xh,\yc) node[above=1mm]{{$a$}};
		
		\draw   (\xk,\ya) to[capacitor,l= {$C_{sst}$}] (\xk,\yc);				
		\draw   (\xl,\ya) to[resistor,l= {$G_{rt}$},-*] (\xl,\yc) node[above=1mm]{{$t$}};
		\draw   (\xm,\ya) to[inductor,l= {$L_{rt}$}] (\xm,\yc);
		
\end{circuitikz}
}
\caption{Equivalent circuit for half-car using $\Pi$-network model.}%
\label{fig:final_coupling_half_car}%
\end{figure*}

The circuit in figure \ref{fig:half_car_coupling} can be simplified. The coupling capacitors can be replaced \cite{zhang2016four} by  two different electrical models: voltage  controlled current sources (VCCS) or capacitors arranged in $\Pi$ network. Figure \ref{fig:coupling_models} shows the layout for both configurations and figure \ref{fig:final_coupling_half_car} shows the simplified circuit using the $\Pi$ network. The values of the inductors and resistors follows the rules on table \ref{tab:tabla1}, while the values of the coupling capacitors are $C_A=\frac{m{l_t}^2+I_G}{l^2}$ and $C_B=\frac{m{l_d}^2+I_G}{l^2}$ and the coupling coefficient is $C_M=\frac{{I_G}-{l_d}{l_t}m}{l^2}$ as reflected in (\ref{eq:m_matrix_halfcar_coupling}).  To facilitate the resolution by applying KCL for a steady state harmonic analysis, the real voltage sources has been transformed into real current sources $i_t$ and $i_d$. This is one the main benefits of using the electromechanical analogy: a plethora of rules, theorems and laws can be applied to simplify the proposed circuits. The resulting matrix equation is showed in (\ref{eq:nueva}).

\begin{equation}
    \begin{bmatrix}
    \vec{Y}_{11} & -C_M\omega j & -b_{sd}-\dfrac{1}{L_{sd}\omega j} & 0 \\
-C_M\omega j & \vec{Y}_{22}  & 0 &-G_{st}-\dfrac{1}{L_{st}\omega j} \\[0.5cm]
-G_{sd}-\dfrac{1}{L_{sd}\omega j} & 0 & \vec{Y}_{33} & 0 \\[0.5cm]
0 & -G_{st}-\dfrac{1}{L_{st}\omega j} & 0 & \vec{Y}_{44}
    \end{bmatrix}
       \begin{bmatrix}
    \vec{U}_a \\[0.5cm] \vec{U}_b\\[0.5cm] \vec{U}_d\\[0.5cm] \vec{U}_t
    \end{bmatrix}
    =
       \begin{bmatrix}
     0\\[0.5cm] 0\\[0.5cm] \vec{I}_{d}\\[0.5cm] \vec{I}_{t}
    \end{bmatrix}
    \label{eq:nueva}
\end{equation}
where

\begin{align*}
  \vec{Y}_{11} &= G_{sd}+C_A\omega j +\dfrac{1}{L_{sd}\omega j}\\ 
    \vec{Y}_{22} &= G_{st}+C_B\omega j+\dfrac{1}{L_{st}\omega j}\\ 
    \vec{Y}_{33} &= C_{ssd}\omega j+G_{sd}+G_{rd}+\dfrac{1}{L_{sd}\omega j}+\dfrac{1}{L_{rd}\omega j}\\ 
    \vec{Y}_{44} &= C_{sst}\omega j + G_{st}+G_{rt}+\dfrac{1}{L_{st}\omega j}+\dfrac{1}{L_{rt}\omega j}\\
    \vec{I}_{d} &= G_{rd}\vec{V}_{d}+\dfrac{1}{L_{rd}\omega j} \vec{V}_{d}\\
    \vec{I}_{t} &= G_{rt}\vec{V}_{t}+\dfrac{1}{L_{rt}\omega j} \vec{V}_{t}\\
\end{align*}

Note again, that gravitational forces (equivalent to DC sources) has been omitted for harmonic analysis. The terms $\vec{V}_d$ and $\vec{V}_t$ represents the voltage source analogue to the velocity of each unsprung mass caused by the road profile.  Also note that in frequency domain $\frac{d}{dt}=j\omega$ and $\int dt = \frac{1}{j\omega}$. It can be observed that mechanical Equations (\ref{eq:m_matrix_halfcar_coupling})-(\ref{eq:equations_Q_half_car}) are the time domain  analogue version of the electrical frequency domain equations in (\ref{eq:nueva}).

\section{Application to a vehicle model with a higher level of complexity}
\begin{figure}%
\centering
\includegraphics[width=0.6\columnwidth]{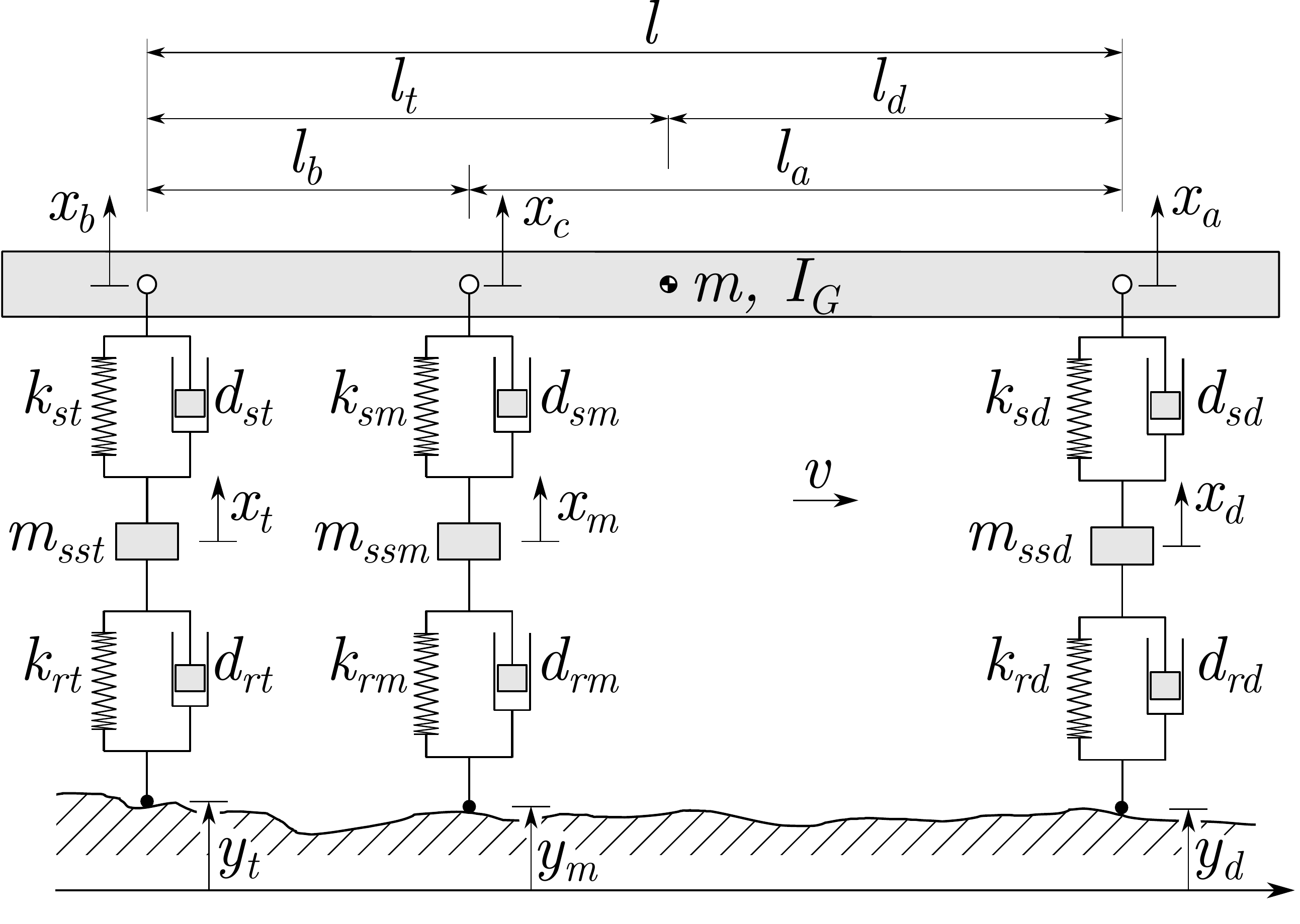}%
\caption{Three-axle vehicle model for vertical dynamics analysis.}%
\label{fig:three_axle}%
\end{figure}

In this section, the system to be analysed is a three-axle vehicle model, which has interest due to the fact that the vertical displacement of the attachment point of the middle axle, $x_c$, is dependent on $x_a$ and $x_b$. In other words, $x_c$ is a dependent coordinate which is written in terms of $x_a$ and $x_b$ as follows
\begin{equation}
x_c=\left({{l_{b}}\,{{x}_{a}}-{l_{b}}\,{{x}_{b}}+{l_{d}}\,{{x}_{b}}+{l_{t}}\,{{x}_{b}}}\right)/l
\end{equation}
where distances $l_a$, $l_b$, $l_d$ and $l_t$ are defined in figure~\ref{fig:three_axle}. In order to keep the text of the manuscript to a reasonable size, the reader is referred to figure~\ref{fig:three_axle} to find the meaning of the mass, damping and stiffness constants of this system.

This system is modeled in terms of the following coordinate vector
\begin{equation}
\vc{x}=\left[\begin{array}{ccccc} {{x}_{a}} & {{x}_{b}} & {{x}_{d}} & {{x}_{t}} & {{x}_{m}} \end{array}\right]^T
\end{equation}

It can be demonstrated with the help of Lagrange equations that the mass, damping and stiffness matrices, according to Equation~\eqref{eq:equ_structure}, are written as follows
\begin{equation}
\vc{m}=\left[\begin{array}{ccccc} m_{11} & m_{12} & 0 & 0 & 0\\ m_{21} & m_{22} & 0 & 0 & 0\\ 0 & 0 & {m_{ssd}} & 0 & 0\\ 0 & 0 & 0 & {m_{sst}} & 0\\ 0 & 0 & 0 & 0 & {m_{ssm}} \end{array}\right]
\label{eq:three_axes_m}
\end{equation}
where $m_{11}=\left(m{l_{t}}^2+I_{G}\right)/{l^2}$, $m_{22}=\left({m\,{{l_{d}}}^2+{I_{G}}}\right)/{l^2}$, and $m_{12}=m_{12}=\left(-{{I_{G}}+{l_{d}}\,{l_{t}}\,m}\right)/{l^2}$.

\begin{equation}
\vc{d}=\left[\begin{array}{ccccc} d_{11} & d_{12} & -{d_{sd}} & 0 & d_{15} \\ d_{21} & d_{22} & 0 & -{d_{st}} & d_{25} \\ -{d_{sd}} & 0 & {d_{rd}}+{d_{sd}} & 0 & 0\\ 0 & -{d_{st}} & 0 & {d_{rt}}+{d_{st}} & 0\\ d_{51} & d_{52} & 0 & 0 & {d_{rm}}+{d_{sm}} \end{array}\right]
\label{eq:three_axex_matrix_b}
\end{equation}
where $d_{11}={d_{sd}}+{{d_{sm}}\,{{l_{b}}}^2}/{l^2}$, $d_{12}=d_{21} = {{d_{sm}}\,{l_{a}}\,{l_{b}}}/{l^2}$, $d_{15}=d_{51} = -{{d_{sm}}\,{l_{b}}}/{l}$, $d_{22}= {d_{st}}+{{d_{sm}}\,{{l_{a}}}^2}/{l^2}$ and $d_{25}=d_{52}=-{{d_{sm}}\,{l_{a}}}/{l}$. 
\begin{equation}
\vc{k}=\left[\begin{array}{ccccc} k_{11} & k_{12} & -{k_{sd}} & 0 & k_{15} \\ k_{21} & k_{22} & 0 & -{k_{st}} & k_{25} \\ -{k_{sd}} & 0 & {k_{rd}}+{k_{sd}} & 0 & 0\\ 0 & -{k_{st}} & 0 & {k_{rt}}+{k_{st}} & 0\\ k_{51} & k_{52} & 0 & 0 & {k_{rm}}+{k_{sm}} \end{array}\right]
\label{eq:three_axex_matrix_k}
\end{equation}
where $k_{11}={k_{sd}}+{{k_{sm}}\,{{l_{b}}}^2}/{l^2}$, $k_{12}=k_{21} = {{k_{sm}}\,{l_{a}}\,{l_{b}}}/{l^2}$, $k_{15}=k_{51} = -{{k_{sm}}\,{l_{b}}}/{l}$, $k_{22}= {k_{st}}+{{k_{sm}}\,{{l_{a}}}^2}/{l^2}$ and $k_{25}=k_{52}=-{{k_{sm}}\,{l_{a}}}/{l}$. 

\begin{figure}%
\centering
\resizebox{0.6\columnwidth}{!}{
\begin{circuitikz}[font=\sffamily, american currents]
\coordinate (m)  at (7, 3.5);
  \coordinate (b)  at (0, 10);

    \draw [font=\normalsize]  (0,0) to[capacitor,l={$\dfrac{m l_d}{l}$},-*] (0,7) node[above=1mm]{\LARGE$b$};
    \draw  [font=\LARGE]  (14,0) to[capacitor,l_={$\dfrac{m l_t}{l}$},-*] (14,7) node[above=1mm]{{\LARGE$a$}};

  \draw  [font=\LARGE]                             (-1,7)      
            to[short]                    (0,7)
            to[capacitor,l_= {$\dfrac{I_G-ml_dl_t}{l^2}$}]    (14,7)
            to[short]                    (15,7);
\draw (-1,0) to[short] (15,0);
\node[circle,fill,label={above:\LARGE$m$}] at (7,3.5) {};

\draw [font=\LARGE] (7,3.5) to [short] (4,2.5) to[short] (4,2.2) to [short] (3.3,2.2) to [short] (4.8,2.2) to [resistor, l={$\dfrac{1}{d_{rm}}$}] (4.8,0);
\draw [font=\LARGE] (3.3,2.2) to [inductor, l_={$\dfrac{1}{k_{rm}}$}] (3.3,0);
\draw [font=\LARGE] (7,3.5) to [capacitor,l= {$m_{ssm}$}]    (7,0);
\draw  [font=\LARGE] (7,3.5) to [short] (10,2.5) to[american current source, l={$Q_{rm}$}, invert] (10,0);

\draw [font=\LARGE] (7,3.5) -- ++ (153.43:2) -- ++ (243.43:0.6) to [resistor, l={$R_{sma}$}] ++(153.43:2) -- ++(63.43:1.2) to [inductor, l={$L_{sma}$}] ++(-26.57:2) -- ++ (243.43:0.6);
\draw (0,7) -- ++(-26.57:3.826);

\draw [font=\LARGE] (7,3.5) -- ++ (26.57:2) -- ++ (116.57:0.6) to [resistor, l={$R_{smb}$}] ++(26.57:2) -- ++(-63.43:1.2) to [inductor, l={$L_{smb}$}] ++(-153.43:2) -- ++ (-243.43:0.6);
\draw (14,7) -- ++(-153.43:3.826);

\draw [font=\LARGE] (5,7) -- ++(90:2) -- ++(0:0.5) -- ++(90:0.6) to [inductor, l={$-L_{smab}$}] ++(0:3) -- ++(-90:1.2) to [resistor, font=\LARGE, l={$-R_{smab}$}] ++(180:3) -- ++(90:0.6);
\draw (9,7) -- ++(90:2) -- ++(180:0.5);
\node[circle,fill,minimum size=5pt, inner sep=-2] at (9,7) {};
\node[circle,fill,minimum size=5pt, inner sep=-2] at (5,7) {};
\end{circuitikz}

}
\caption{Electrical analogue of the three-axle vehicle model (zoomed in the coupling area of figure \ref{fig:final_coupling_half_car}). 
$L_{sma}=\dfrac{l}{k_{sm}l_a}$, $L_{smb}=\dfrac{l}{k_{sm}l_b}$, $R_{sma}=\dfrac{l}{R_{sm}l_a}$, $R_{smb}=\dfrac{l}{k_{sm}l_b}$, $L_{smab}=\dfrac{l}{k_{sm}l_al_b}$ and $R_{smab}=\dfrac{l}{R_{sm}l_al_b}$}%
\label{fig:new_three_wheels_analog}%
\end{figure}
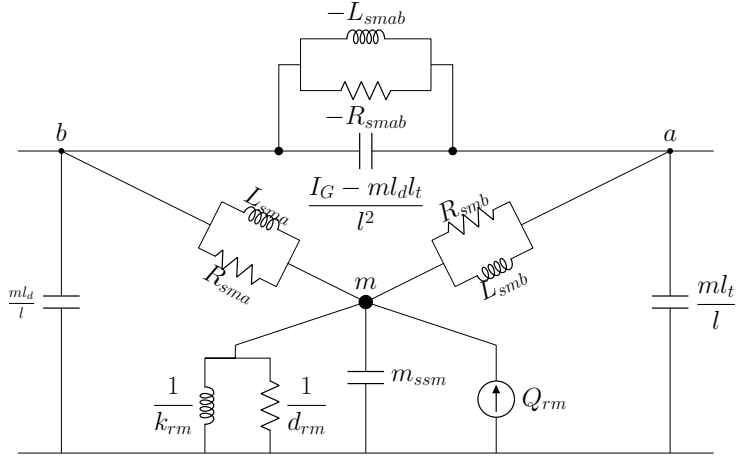

The generalized force vector is written as follows:
\begin{equation}
\vc{Q}(t)=\left[\begin{array}{c} -\dfrac{g\,{l_{t}}\,m}{l}\\ -\dfrac{g\,{l_{d}}\,m}{l}\\ -g\,{m_{ssd}}\\ -g\,{m_{sst}}\\ -g\,{m_{ssm}} \end{array}\right]
+\left[\begin{array}{c} 0\\ 0\\ {d_{rd}}\,{\dot{y}_{d}}(t)+{k_{rd}}\,{{y}_{d}}(t)\\ {d_{rt}}\,{\dot{y}_{t}}(t)+{k_{rt}}\,{{y}_{t}}(t)\\ {d_{rm}}\,{\dot{y}_{m}}(t)+{k_{rm}}\,{{y}_{m}} (t)\end{array}\right]\label{eq:vectorQthreeaxle}
\end{equation}
where ${{y}_{d}}$, ${{y}_{m}}$ and ${{y}_{t}}$ are the front, middle and rear vertical displacements of the wheel and ground contact points. 
It should be noted that matrix (\ref{eq:three_axes_m}) is very similar to (\ref{eq:m_matrix_halfcar_coupling}) but with a new element in the diagonal that accounts for the mass $m_{ssm}$. However, matrix equations (\ref{eq:three_axex_matrix_b}) and (\ref{eq:three_axex_matrix_k}) differ to those of a half-car. Now, new elements arise in positions $12$ (and symmetrically in $21$). Also, elements $11$ and $22$ change their value. The equivalent electrical model is shown in figure \ref{fig:new_three_wheels_analog}.  As a result, the half-car model is extended by adding a new node, $m$, (due to the new axle) with new elements connecting node $a$ and $b$ with $m$. Also a new branch connecting $a$ and $b$ is observed. Interestingly enough, this branch has elements with negative values. Finally, the node $m$ is linked to ground through the expected elements in a similar fashion as the other two wheels. The road profile of the third wheel is modelled through a new voltage source $v_m$. 

The new matrix equation reads as follows:
\begin{equation}
    \begin{bmatrix}
    \vec{Y}_{11} & \vec{Y}_{12} & \vec{Y}_{13} & 0 & \vec{Y}_{15}\\
\vec{Y}_{21} & \vec{Y}_{22}  & 0 &\vec{Y}_{24} & \vec{Y}_{25}\\
\vec{Y}_{31} & 0 & \vec{Y}_{33} & 0 & 0\\
0 & \vec{Y}_{42} & 0 & \vec{Y}_{44} & 0\\
\vec{Y}_{51} & \vec{Y}_{52} & 0 &  0 & \vec{Y}_{55} 
    \end{bmatrix}
       \begin{bmatrix}
    \vec{U}_a \\ \vec{U}_b\\ \vec{U}_d\\ \vec{U}_t \\ \vec{U}_m
    \end{bmatrix}
    =
       \begin{bmatrix}
     0\\ 0\\ \vec{I}_{d}\\ \vec{I}_{t} \\ \vec{I}_m
    \end{bmatrix}
    \label{eq:nueva2}
\end{equation}

where

\begin{align*}
    \vec{Y}_{11}&=G_{sd}+C_A\omega j +\dfrac{1}{L_{sd}\omega j}+\left(G_{sm}+\dfrac{1}{L_{sm}\omega j}\right)\dfrac{l_{b}^2}{l^2}&\\ \nonumber
    \vec{Y}_{22}&=G_{st}+C_B\omega j+\dfrac{1}{L_{st}\omega j}+\left(G_{sm}+\dfrac{1}{L_{sm}\omega j}\right)\dfrac{l_{a}^2}{l^2}\\ \nonumber
    \vec{Y}_{33}&=C_{ssd}\omega j+G_{sd}+G_{rd}+\dfrac{1}{L_{sd}\omega j}+\dfrac{1}{L_{rd}\omega j}\\ \nonumber
    \vec{Y}_{44}&=C_{sst}\omega j + G_{st}+G_{rt}+\dfrac{1}{L_{st}\omega j}+\dfrac{1}{L_{rt}\omega j}\\\nonumber
    \vec{Y}_{55}&=C_{ssm}\omega j + G_{sm}+G_{rm}+\dfrac{1}{L_{sm}\omega j}+\dfrac{1}{L_{rm}\omega j}\\\nonumber
    \vec{Y}_{12}&=\vec{Y}_{21}=-C_{M}\omega j + \left(G_{sm}+\dfrac{1}{L_{sm}\omega j}\right)\dfrac{l_{a}l_{b}}{l^2}\\\nonumber
    \vec{Y}_{13}&=\vec{Y}_{31}-b_{sd}-\dfrac{1}{L_{sd}\omega j}\\\nonumber
    \vec{Y}_{15}&=\vec{Y}_{51}=-\left(G_{sm}+\dfrac{1}{L_{sm}\omega j}\right)\dfrac{l_{b}}{l}\\\nonumber
    \vec{Y}_{24}&=\vec{Y}_{42}=-G_{st}-\dfrac{1}{L_{st}\omega j} \\\nonumber
    \vec{Y}_{25}&=\vec{Y}_{52}=-\left(G_{sm}+\dfrac{1}{L_{sm}\omega j}\right)\dfrac{l_{a}}{l}\\\nonumber
    \vec{I}_{d}&=G_{rd}\vec{V}_{d}+\dfrac{1}{L_{rd}\omega j} \vec{V}_{d},\quad  \vec{I}_{t}=G_{rt}\vec{V}_{t}+\dfrac{1}{L_{rt}\omega j} \vec{V}_{t} \\
    \vec{I}_{m}&=G_{rm}\vec{V}_{m}+\dfrac{1}{L_{rm}\omega j} \vec{V}_{m}
\end{align*}

\section{Numerical results. Validation of electromechanical analogues}
To show the benefits of the presented analogy, a numerical example of an electrical analogue of a three-axle vehicle model in steady state harmonic vibration is solved. The model is inspired in the real three-axle heavy truck modeled by Wang et al.~\cite{Wang2020}. The constants of the lumped parameter truck model are summarized in Table~\ref{tab:three_axle}.

\begin{table}[ht]%
\centering

\begin{tabular}{lcrr}
\toprule
\textbf{Parameter} & \textbf{Symbol} & \textbf{Value} & \textbf{Units}\\ \midrule
mass & $m$ & 22,000 & kg\\
pitch moment of inertia & $I_G$ & 21,000 &kgm$^2$\\
front axle unsprung mass & $m_{ssd}$ & 900 & kg\\
middle axle unsprung mass & $m_{ssm}$ & 1,400 & kg\\
rear axle unsprung mass & $m_{sst}$ & 1,400 & kg\\ \hline
front axle suspension stiffness & $k_{sd}$ & 610,000 & N/m\\
middle axle suspension stiffness & $k_{sm}$ & 2,600,000 & N/m\\
rear axle suspension stiffness & $k_{st}$ & 2,600,000 & N/m\\
front axle suspension damping & $d_{sd}$ & 15,400 &Ns/m\\
middle axle suspension damping & $d_{sm}$ & 15,400 &Ns/m\\
rear axle suspension damping & $d_{st}$ & 15,400 &Ns/m\\ \hline
front axle tyre stiffness & $k_{rd}$ & 1,360,000 &N/m\\
middle axle tyre stiffness & $k_{rm}$ & 5,430,000 &N/m\\
rear axle tyre stiffness & $k_{rt}$ & 5,430,000 &N/m\\
front axle tyre damping & $d_{rd}$ & 150 &Ns/m\\
middle axle tyre damping & $d_{rm}$ & 150 &Ns/m\\
rear axle tyre damping & $d_{rt}$ & 150 &Ns/m\\ \hline
rear to front axle distance & $l$ & 6.15 &m\\
truck center of mass to front axle distance & $l_d$ & 4.44 &m\\
truck center of mass to rear axle distance & $l_t$ & 1.71 &m\\
middle to front axle distance & $l_a$ & 4.80 &m\\
middle to rear axle distance & $l_b$ & 1.35 &m\\ 
\bottomrule
\end{tabular}
\caption{Constants of the lumped parameter three axle vehicle model.}
\label{tab:three_axle}
\end{table} 

The vehicle model is assumed to travel at a constant forward velocity, $v =60$ km/h, on a road with a harmonic unevenness characterized by an amplitude, $Y= 5$ cm and a wave length, $\lambda= 2$ m. This way, the displacements of the front, middle and rear tyre-ground contact points can be modelled as follows:
\begin{equation}
\begin{aligned}
&y_d(t) = Y\sin\left({2\pi v t}/{\lambda}\right)\\ 
&y_m(t) = Y\sin\left({2\pi v t}/{\lambda}-\phi_m\right)\\
&y_t(t) = Y\sin\left({2\pi v t}/{\lambda}-\phi_t\right) 
\end{aligned}    
\end{equation}

where $\phi_m = 2\pi{l_a}/{\lambda}$ and $\phi_t = 2\pi{l}/{\lambda}$ are phase shifts. 

To solve the electrical analogue in harmonic steady state analysis, we must first perform the complex phasor representation of the variables presented in table \ref{tab:three_axle} and then insert them into equation (\ref{eq:nueva2}) to solve the linear matrix system.  Table \ref{tab:results_3_axle} shows the complex phasor values of  elements in  (\ref{eq:nueva2}) for this example. The vector current is also presented in complex form.  The unknowns $  \vec{U} = [\vec{U}_a,\vec{U}_b,\vec{U}_d,\vec{U}_t,\vec{U}_m]^T$ are easily obtained by performing the inverse of the admittance matrix and then multiply the current vector $\vec{I}$. The result is
\begin{gather}
  \begin{bmatrix}
           \vec{U}_{a} \\
           \vec{U}_{b} \\
           \vec{U}_{d} \\
           \vec{U}_{t} \\
           \vec{U}_{m}
         \end{bmatrix}
         =
         \begin{bmatrix}
   0.0754 + 0.5293i\\
  -0.0987 + 0.0033i\\
  -0.8114 - 2.1287i\\
   1.7945 - 1.4511i\\
  -2.1784 - 0.9272i\\
         \end{bmatrix}
         =          \begin{bmatrix}
    0.5346\times e^{81.90 i}\\
    0.0988\times e^{178.11 i}\\
    2.2781\times e^{-110.87 i}\\
    2.3078\times e^{-38.96 i}\\
    2.3675\times e^{-156.94 i}\\
         \end{bmatrix}
         \label{eq:result_rms_3-axle}
  \end{gather}
The result in (\ref{eq:result_rms_3-axle}) are complex values representing the voltage (velocities) in the electrical (mechanical) circuit. The norm of the complex phasor gives the RMS amplitude of for each harmonic sinusoidal waveform. The time domain expressions are
\begin{equation}\begin{split}
    u_a(t)&=\sqrt{2}\;0.5346 \sin(\omega t+81.90)\\ 
    u_b(t)&=\sqrt{2}\;0.0988 \sin(\omega t+178.11)\\ 
    u_d(t)&=\sqrt{2}\;2.2781 \sin(\omega t-110.87)\\
    u_t(t)&=\sqrt{2}\;2.3078 \sin(\omega t-38.96)\\
    u_m(t)&=\sqrt{2}\;2.3675 \sin(\omega t-156.94) \label{eq:rms_values}
\end{split}\end{equation}
in Volt (meter/second) for the electrical (mechanical) circuit. The phase angle are expressed in degrees.

\begin{table}[]
\resizebox{\columnwidth}{!}{%
\begin{tabular}{llrrrrr}
\toprule
& &\multicolumn{1}{c}{1} & \multicolumn{1}{c}{2} & \multicolumn{1}{c}{3} & \multicolumn{1}{c}{4} & \multicolumn{1}{c}{5}\\ \cmidrule(l){3-3}\cmidrule(l){4-4}\cmidrule(l){5-5}\cmidrule(l){6-6}\cmidrule(l){7-7}
\multirow{5}{*}[-1em]{\begin{sideways}  $[\vec{Y}]$ matrix\end{sideways}}\\
&\multicolumn{1}{r|}{1} &0.1614 + 1.0408\textit{j} &  0.0264 + 1.9365\textit{j} & -0.1540 + 0.1165\textit{j}  & 0 &  -0.0338 + 0.1090\textit{j}\\
&\multicolumn{1}{r|}{2}&   0.0264 + 1.9365\textit{j} &  0.2478 + 5.4956\textit{j} &  0 & -0.1540 + 0.4966\textit{j} & -0.1202 + 0.3876\textit{j}\\
&\multicolumn{1}{r|}{3}&  -0.1540 + 0.1165\textit{j} &  0  & 0.1555 + 0.0950\textit{j} &  0 &  0\\
&\multicolumn{1}{r|}{4}&   0 & -0.1540 + 0.4966\textit{j}  & 0 &  0.1555 $-$ 0.8006\textit{j} &  0\\
&\multicolumn{1}{r|}{5}&  -0.0338 + 0.1090\textit{j} & -0.1202 + 0.3876\textit{j}  & 0 &  0 &  0.1555 $-$ 0.8006\textit{j}\\ \hline
 &\multicolumn{1}{r|}{$\vec{I}$} & 0&0&   0.0028 - 0.4808\textit{j} &  -0.8691 - 1.7118\textit{j}&  -1.1307 + 1.5515\textit{j}\\
\bottomrule
\end{tabular}%
}
\caption{Complex phasor  data in (\ref{eq:nueva2}) for 3-axle numerical example. All values are multiplied by $10^{-5}$ }
\label{tab:results_3_axle}
\end{table}

\begin{figure}[htbp]%
\centering
\includegraphics[trim=0 0 0 0, clip,width=\linewidth]{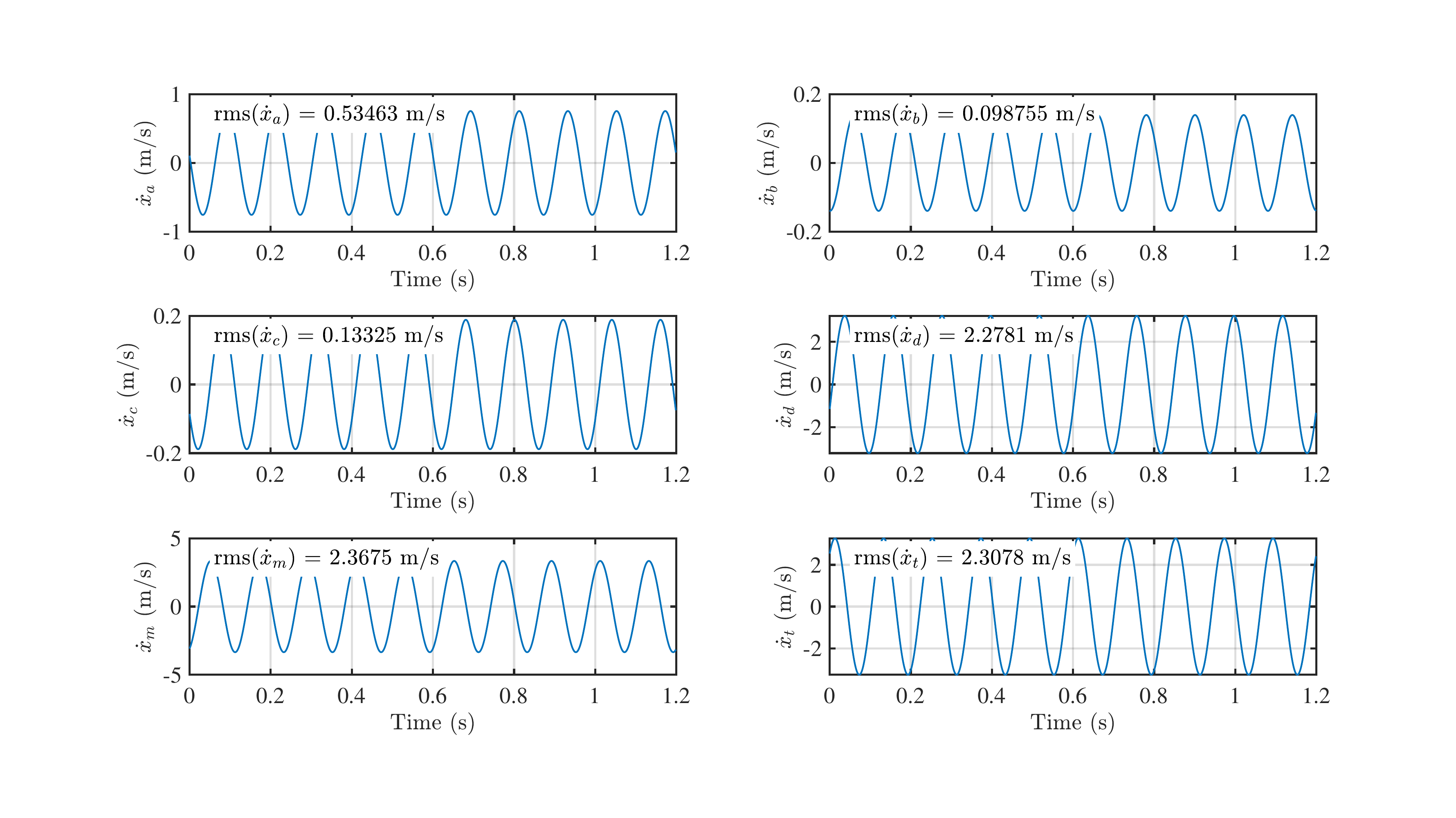}%
\caption{Steady state vertical oscillation velocities of the characteristic points ($a$, $b$, $c$, $d$, $m$, and $t$) of the three-axle vehicle model.}%
\label{fig:velocity}%
\end{figure}

In order to validate the previous results, the steady state harmonic vibration of the three-axle vehicle model is studied with the help of the inverse Fourier transform. As it is well known, the steady state oscillation velocity vector can be obtained as follows:
\begin{equation}
\vc{\dot{x}}(t) = \int_{-\infty}^{\infty}{j\omega\vc{H}(\omega)\vc{Q_h}(\omega)\mathrm{e}^{j\omega t}\mathrm{d}\omega}
\label{eq:dominio_frec2}
\end{equation}
where $\vc{H}(\omega) = \left(-\omega^2\vc{m}+j\omega\vc{d}+\vc{k}\right)^{-1}$ is the frequency response matrix function and $\vc{Q_h}(\omega)$ is the Fourier transform of the harmonic part of the excitation, which is obtained as follows
\begin{equation}
\vc{Q_h}(\omega)=\frac{1}{2\pi}\int_{-\infty}^{\infty}{\left[\begin{array}{c} 0\\ 0\\ {d_{rd}}\,{\dot{y}_{d}}(t)+{k_{rd}}\,{{y}_{d}}(t)\\ {d_{rt}}\,{\dot{y}_{t}}(t)+{k_{rt}}\,{{y}_{t}}(t)\\ {d_{rm}}\,{\dot{y}_{m}}(t)+{k_{rm}}\,{{y}_{m}} (t)\end{array}\right]\mathrm{e}^{-j\omega t}\mathrm{d}t}
\label{eq:qsubh}
\end{equation}
Note that the constant part of the excitation vector appearing in Equation~\eqref{eq:vectorQthreeaxle} has been ignored for this analysis since it does not affect the velocity in steady state. The velocities in Equation~\eqref{eq:dominio_frec2} have been solved with the help of the \texttt{ifft} subroutine, while $\vc{Q_h}(\omega)$ is computed with the \texttt{fft} sobroutine, both from Matlab. For the solution, a total of ten oscillations with 1024 time points have been simulated. This gave a sampling frequency of 833.33 Hz, for an excitation frequency of 8.33 Hz.  Finally, the velocity $\dot{x}_c(t)$ is obtained from the velocities of points $a$ and $b$ as follows:
\begin{equation}
\dot{x}_c=\left({{l_{b}}\,{\dot{x}_{a}}-{l_{b}}\,{\dot{x}_{b}}+{l_{d}}\,{\dot{x}_{b}}+{l_{t}}\,{\dot{x}_{b}}}\right)/l
\end{equation}
The velocities obtained for the characteristic points ($a$, $b$, $c$, $d$, $m$, and $t$) of the three-axle vehicle model are shown in Figure~\ref{fig:velocity} together with the rms values of the velocity signals for comparison against the results of Equation~\eqref{eq:rms_values}. As it can be seen, the rms values exactly coindice with those of the electric analogue circuit.

\section{Conclusions}
The use of electric analogues of vehicle models is a topic of interest and has been studied previously in the literature. In particular, the quarter car model has been studied in several publications. Such electric analogues have been successfully used for tunning controllers of active suspension elements. Nevertheless,  more complex models such as the half car vehicle model  lacks of a comprehensive electrical analogue. This work attempts to shed some light on this subject by providing the electrical analogue of the half car model as well as its closely related, but more complicated model, three-axle vehicle. Therefore, new and non-previously disclosed electrical analogues for two moderately complex vehicle models have been described. 

The inertial coupling present in such mechanical systems due to the vehicle main frame , has been identified and modelled for the first time by an electrostatic capacitor coupling in its force-current analogue. Only pure displacement coordinates has been used instead of mixing angles and displacemets variables. The capacitance values of the coupled capacitors are dependent on the mass and moment of inertia of
the vehicle main frame, and on the distance of each wheel to the center of gravity of the main frame.
To deal with the coupling capacitors, its equivalent  $\Pi$ network has been used in order to simplify the resulting equations. Transformation between real voltage and current sources has been also used to facilitate the application of KCL's.

Finally,  numerical results have been obtained for a three-axle vehicle model in steady state harmonic vibration conditions both by using the electrical equivalent circuit and the mechanical model as an example of the utility of this method. As expected, both approaches leaded to the same results by sharing the same time and frequency domain equations.

The presented methodology  would be of interest for further studies where coupled electric and mechanical systems are available. The design of control strategies of active suspension systems is a classic example. Furthermore, the proposed analogy opens up new possibilities for the  application of some well-known tools in Circuit Theory analysis.

\section*{Acknowledgment}

This research has been supported by the Spanish Ministry of Science, Innovation and Universities 
under the programme \textit{Proyectos de I+D de Generacion de Conocimiento} of the 
R+D+I system with grant number PGC2018-098813-B-C33.


\bibliographystyle{elsarticle-num}
\bibliography{paper_analogies}

%








\end{document}